\documentstyle[aps]{revtex}
\begin{document}
\author{M. Castagnino, R. Laura.}
\address{CONICET-Instituto de F\'{i}sica, Rosario, Argentina.}
\author{O. Lombardi}
\address{CONICET-IEC Universidad de Quilmes, Buenos Aires, Argentina.}
\title{Decoherence according to Environment and Self Induced Decoherences.}
\maketitle

\begin{abstract}
A generalized decoherence formalism that can be used both in open (using
Environment Induced Decoherence-EID) and closed (using Self Induced
Decoherence-SID) quantum systems is sketched.
\end{abstract}

\section{Introduction}

Quantum mechanics is essentially characterized by the superposition
principle and the phenomenon of interference. Therefore, any attempt to
explain how classicality emerges from quantum behavior must include two
elements: a process through which interference vanishes, and a
superselection rule that precludes superpositions. Decoherence is the
process that eliminates interference and leads to the rule that selects the
candidates for classical states.

Schematically, three periods can be identified in the development of the
theory:

{\bf Closed systems period} (van Kampen, van Hove, Daneri {\it et al.} \cite
{vK}). In order to understand how classical macroscopic features arise from
quantum microscopic behavior, ''gross'' observables are defined. The states
that are indistinguishable for a macroscopic observer are described by the
same coarse-grained state $\rho _{G}(t)$. When the evolution of $\rho
_{G}(t) $ (or of the expectation value of the gross observables) is studied,
it is proved that $\rho _{G}(t)$ reaches equilibrium in a relaxation time $%
t_{R}$; therefore, $\rho _{G}(t)$ decoheres in its own eigenbasis after a
decoherence time $t_{D}=t_{R}$. The main problem of this period was the fact
that $t_{R}$ turned out to be too long to account for experimental data (see 
\cite{Omnes}).

{\bf Open systems period}. An open system $S$ is considered in interaction
with its environment $E$, and the evolution of the reduced state $\rho
_{S}(t)=Tr_{E}\rho (t)$ is studied. The so called ''environment induced
decoherence'' (EID) (Zeh and Zurek \cite{Zeh}) proves that, since the
interference terms of $\rho _{S}(t)$ rapidly vanish, $\rho _{S}(t)$
decoheres in an adequate pointer basis after a short decoherence time $%
t_{D}=t_{DS}$. This result solved the main problem of the first period.

{\bf Closed and open systems period.} Although at present EID is still
considered the ''orthodoxy'' in the subject (\cite{Omnes}), other approaches
have been proposed to face the problems of EID, in particular, the
closed-system problem (Diosi, Milburn, Penrose, Casati and Chirikov, Adler 
\cite{Penrose}). Some of these methods are clearly ''non-dissipative'' (Ford
and O'Connell \cite{O'Connel}), that is, not based on the dissipation of
energy from the system to the environment. Among them, we have developed the
self-induced decoherence (SID) approach, according to which a closed quantum
system with continuous spectrum may decohere by destructive interference,
and reaches a final state where the classical limit can be obtained (\cite
{PRA-2000}).

In spite of the fact that the theories of decoherence in closed and open
systems coexist in the third period, in the literature both kinds of
approaches are usually conceived as antagonistic scenarios for decoherence
or even as formalism dealing with different physical phenomena (\cite{Max}).
In this paper we will argue that this is not the case; on the contrary, the
two kinds of theories can be subsumed under a more general theoretical
framework that shows their complementary character.

\section{Observables, mean values, and weak limits}

As emphasized by Omn\'{e}s (\cite{Omnes}), decoherence is a particular case
of the phenomenon of irreversibility, which leads to the following problem.
Since the quantum state $\rho (t)$ evolves unitarily, it cannot reach a
final equilibrium state for $t\rightarrow \infty $. Therefore, if the
non-unitary evolution towards equilibrium is to be accounted for, a further
element has to be added, which consists in the splitting of the maximal
information about the system into a relevant part and an irrelevant part:
whereas the irrelevant part is discarded, the relevant part reaches a final
equilibrium situation. In some cases, the irrelevant part can be conceived
as an environment, but this is not always necessary. When this idea is
rephrased in operator language, the phenomenon of decoherence can be
explained in three general steps:

1.- The set ${\cal O}$ of relevant observables is defined. The observables $%
O_{R}\in {\cal O}$ restrict the maximal information given by the set of all
the possible observables of the system.

2.- The expectation value $\langle O_{R}\rangle _{\rho (t)}$ is computed,
for any $O_{R}\in {\cal O}$.

3.- It is proved that $\langle O_{R}\rangle _{\rho (t)}$ reaches a final
equilibrium value expressed by the limit of $\langle O_{R}\rangle _{\rho (t)}
$ or the weak limit of $\rho (t)$: 
\[
\lim_{t\rightarrow \infty }\langle O_{R}\rangle _{\rho (t)}=\langle
O_{R}\rangle _{\rho _{*}}\qquad or\qquad W\lim_{t\rightarrow \infty }\rho
(t)=\rho _{*}
\]
These three stages can be found in both open and closed systems; in
particular, we will find them in the example of section IV.

When decoherence in closed and open systems is understood in the context of
this common general framework, the relationship between both cases can be
studied. Let us consider a large closed system $U$ (''the universe'') that
can be split into many subsystems $S_{i}$ such that $U=\bigcup_{i}S_{i}$,
each one of them with its own environment $E_{i}=U\backslash S_{i}$. Let $%
\rho (t)$ be the state of $U$ and $\rho _{S_{i}}(t)$ the state of $S_{i}$.
Under very general conditions, it can be proved that:

(i) The decoherence of the closed system $U$ yields the decoherence of all
its open subsystems $S_{i}$ in interaction with their environments $E_{i}$.
In other words, if $\rho (t)$ reaches an equilibrium state $\rho _{*}$
(according to SID), which is diagonal in its pointer basis, then all the $%
\rho _{S_{i}}(t)$ also reach equilibrium states $\rho _{S_{i*}}$ (according
to EID), each one diagonal in its own pointer basis.

(ii) The decoherence of all the open subsystems $S_{i}$ in interaction with
their environments $E_{i}$ yields the decoherence of the closed system $U$.
In other words, if all the $\rho _{S_{i}}(t)$ reach equilibrium states $\rho
_{S_{i*}}$ (according to EID), each one diagonal in its pointer basis, then $%
\rho (t)$ reaches an equilibrium state $\rho _{*}$ (according to SID), which
is diagonal in its own pointer basis.

Of course, each process has its own decoherence time: these times may be
different and, eventually, some of them may be infinite (in this case,
decoherence is merely theoretical). We will develop this general idea in
mathematical terms elsewhere. Nevertheless, this shows that the formalisms
for open and closed systems are complementary, and both cooperate in the
understanding of the same physical reality.

\section{Decoherence time, triviality, and complexity}

For open systems, the decoherence time $t_{DS}$ obtained in several models
studied by EID (precisely, the characteristic time of the diagonalization of 
$\rho _{S}(t)$ in a pointer basis obtained by means of the predictability
sieve criterion) is given by eq.(47) of \cite{3A} or by eq.(3.136) of \cite
{Ex}.

For closed systems, in paper \cite{DT} we have found the decoherence time $%
t_{DU}$ by the SID method, and we have explained that such a time is {\it %
infinite} when the Hamiltonian and the initial conditions are {\it trivial},
(i.e., just with real poles in the analytical continuation of the resolvent
and the initial conditions). In the Appendix B of that paper we, have showed
how the method can describe a two-times evolution (that can be easily
generalized to an n-times evolution). Let us rephrase that appendix: by
choosing $V^{(1)}$ as representing the interaction between a proper system $S
$ and its environment $E$, $V^{(2)}$ as representing the interaction of the
parts of the environment $E$ among themselves, and $V^{(1)}(\omega ,\omega
^{\prime })\gg V^{(2)}(\omega ,\omega ^{\prime })$, we will find that the
first interaction gives the decoherence time $t_{DS}$ of the proper system $S
$ and the second interaction gives the decoherence time $t_{DU}$ of the
whole system $U$. These  interactions are what preclude triviality and
introduce complexity. Moreover, as $V^{(1)}(\omega ,\omega ^{\prime })\gg
V^{(2)}(\omega ,\omega ^{\prime })$ and $t_{D}\sim \frac{\hbar }{V}$, we
obtain $t_{DS}\ll $ $t_{DU}.$ As an example (see \cite{DT} for details), if
we assume a microscopic self-environment interactions $V^{(2)}\approx 1$ $eV$%
, and a macroscopic oscillator-environment interaction $V^{(1)}\approx
10^{23}eV$, we have $t_{DU}\approx 10^{-15}s$ and $t_{DS}\approx
10^{-37}s-10^{-39}s$ . As a consequence, we certainly obtain $t_{DS}\ll $ $%
t_{DU}.$

\section{A well known model}

Let us consider a spin system $S_{0}$ (with Hilbert space ${\cal H}_{0}$)
and a set of $N$ spin systems $S_{i}$ (with Hilbert spaces ${\cal H}_{i}$)
coupled by the Hamiltonian (see \cite{Max}) 
\begin{equation}
H=H_{SE}=\frac{1}{2}(|0\rangle \langle 0|-|1\rangle \langle
1|)\sum_{i=0}^{N}g_{i}(|\uparrow _{i}\rangle \langle \uparrow
_{i}|-|\downarrow _{i}\rangle \langle \downarrow _{i}|)\bigotimes_{j\neq
i}^{N}I_{j}  \label{7}
\end{equation}
We will also suppose that the free Hamiltonians are $H_{S}=H_{E}=0$.

Let us consider a pure state $|\psi _{0}\rangle =(a|0\rangle +b|1\rangle
)\bigotimes_{i}^{N}(\alpha _{i}|\uparrow _{i}\rangle +\beta _{i}|\downarrow
_{i}\rangle )$, where $\alpha _{i}$ and $\beta _{i}$ are aleatory
coefficients such that $|\alpha _{i}|^{2}+|\beta _{i}|^{2}=1$. The state $%
|\psi _{0}\rangle $ evolves as $|\psi (t)\rangle =a|0\rangle |{\cal E}%
_{0}(t)\rangle +b|1\rangle |{\cal E}_{1}(t)\rangle $, where $|{\cal E}%
_{0}(t)\rangle =|{\cal E}_{1}(-t)\rangle =\bigotimes_{i}^{N}(\alpha
_{i}e^{ig_{i}t/2}|\uparrow _{i}\rangle +\beta _{i}e^{-ig_{i}t/2}|\downarrow
_{i}\rangle )$, and its matrix version is $\rho (t)=|\psi (t)\rangle \langle
\psi (t)|$. Now we can develop the three steps introduced in Section II in
the case of this model.

{\bf Step 1}: We will consider the observables $O\in {\cal O}$ such that 
\[
O=(s_{00}|0\rangle \langle 0|+s_{01}|0\rangle \langle 1|+s_{10}|1\rangle
\langle 0|+s_{11}|1\rangle \langle 1|)\bigotimes_{i=1}^{N}(\epsilon
_{\uparrow \uparrow }^{(i)}|\uparrow _{i}\rangle \langle \uparrow
_{i}|+\epsilon _{\downarrow \downarrow }^{(i)}|\downarrow _{i}\rangle
\langle \downarrow _{i}|+\epsilon _{\downarrow \uparrow }^{(i)}|\downarrow
_{i}\rangle \langle \uparrow _{i}|+\epsilon _{\uparrow \downarrow
}^{(i)}|\uparrow _{i}\rangle \langle \downarrow _{i}|) 
\]
where $s_{00},s_{11},\epsilon _{\uparrow \uparrow }^{(i)},\epsilon
_{\downarrow \downarrow }^{(i)}$ are real numbers and $s_{01}=s_{10}^{*},$ $%
\epsilon _{\uparrow \downarrow }^{(i)}=\epsilon _{\downarrow \uparrow
}^{(i)*}$ are complex numbers. These observables are not completely general,
but they are the relevant observables for this case (see \cite{Max} eq.(18)).

{\bf Stage 2}: The expectation value of $O$ in the state $|\psi (t)\rangle $
reads 
\[
\langle O\rangle _{\psi (t)}=(|a|^{2}s_{00}+|b|^{2}s_{11})\Gamma _{0}(t)+2%
%TCIMACRO{\func{Re}}
%BeginExpansion
\mathop{\rm Re}%
%EndExpansion
[ab^{*}s_{10}\Gamma _{1}(t)] 
\]
where 
\begin{equation}
\Gamma _{0}(t)=\prod_{i=1}^{N}[|\alpha _{i}|^{2}\epsilon _{\uparrow \uparrow
}^{(i)}+|\beta _{i}|^{2}\epsilon _{\downarrow \downarrow }^{(i)}+\alpha
_{i}{}^{*}\beta _{i}\epsilon _{\uparrow \downarrow
}^{(i)}e^{-ig_{i}t}+(\alpha _{i}{}^{*}\beta _{i}\epsilon _{\uparrow
\downarrow }^{(i)})^{*}e^{ig_{i}t}]  \label{M.1}
\end{equation}
\begin{equation}
\Gamma _{1}(t)=\prod_{i=1}^{N}[|\alpha _{i}|^{2}\epsilon _{\uparrow \uparrow
}^{(i)}e^{ig_{i}t}+|\beta _{i}|^{2}\epsilon _{\downarrow \downarrow
}^{(i)}e^{-ig_{i}t}+\alpha _{i}{}^{*}\beta _{i}\epsilon _{\uparrow
\downarrow }^{(i)}+(\alpha _{i}{}^{*}\beta _{i}\epsilon _{\uparrow
\downarrow }^{(i)})^{*}]  \label{M.2}
\end{equation}

Let us consider two special cases of observables:

(a) If $\epsilon _{\uparrow \uparrow }^{(i)}=\epsilon _{\downarrow
\downarrow }^{(i)}=1,$ $\epsilon _{\uparrow \downarrow }^{(i)}=0$, we are in
the typical case of EID. In this case, 
\[
\text{ }O_{S_{0}}=(s_{00}|0\rangle \langle 0|+s_{01}|0\rangle \langle
1|+s_{10}|1\rangle \langle 0|+s_{11}|1\rangle \langle
1|)\bigotimes_{i=0}^{N}I_{i}
\]
The expectation value of these observables is given by 
\begin{equation}
\langle O_{S}\rangle _{\psi (t)}=|\alpha _{i}|^{2}s_{00}+|\beta
_{i}|^{2}s_{01}+%
%TCIMACRO{\func{Re}}
%BeginExpansion
\mathop{\rm Re}%
%EndExpansion
[ab^{*}s_{10}r(t)]  \label{16}
\end{equation}
where $r(t)=\langle {\cal E}_{1}(t)\rangle |{\cal E}_{0}(t)\rangle $ and 
\begin{equation}
|r(t)|^{2}=\prod_{i=1}^{N}(|\alpha _{i}|^{4}+|\beta _{i}|^{4}+2|\alpha
_{i}|^{2}|\beta _{i}|^{2}\cos 2g_{i}t)  \label{r}
\end{equation}
These are the observables that ''observe'' only the spin system  $S_{0}$.

(b) But we can also define the observables that ''observe'' just {\it one}
spin system  $S_{j}$ of the environment: 
\[
O_{S_{j}}=I_{S_{0}}\otimes O_{j}\bigotimes_{i\neq j}I_{S_{i}}
\]
with 
\[
O_{j}=\epsilon _{\uparrow \uparrow }^{(j)}|\uparrow _{j}\rangle \langle
\uparrow _{j}|+\epsilon _{\downarrow \downarrow }^{(j)}|\downarrow
_{j}\rangle \langle \downarrow _{j}|+\epsilon _{\downarrow \uparrow
}^{(j)}|\downarrow _{j}\rangle \langle \uparrow _{j}|+\epsilon _{\uparrow
\downarrow }^{(j)}|\uparrow _{j}\rangle \langle \downarrow _{j}|
\]
where $\epsilon _{\uparrow \uparrow }^{(j)},\epsilon _{\downarrow \downarrow
}^{(j)},$ $\epsilon _{\uparrow \downarrow }^{(j)}$ are now generic. The
expectation value of these observables is given by 
\[
\langle O_{S_{j}}\rangle _{\psi (t)}=\langle \psi (t)|O_{S_{j}}|\psi
(t)\rangle =|a|^{2}(|\alpha _{j}|^{2}\epsilon _{\uparrow \uparrow
}^{(j)}+|\beta _{j}|^{2}\epsilon _{\downarrow \downarrow }^{(j)}+\alpha
_{j}\beta _{j}^{*}\epsilon _{\uparrow \downarrow }^{(j)}e^{-ig_{j}t}+\alpha
_{j}^{*}\beta _{j}\epsilon _{\downarrow \uparrow }^{(j)}e^{ig_{j}t})+
\]
\begin{equation}
|b|^{2}(|\alpha _{j}|^{2}\epsilon _{\uparrow \uparrow }^{(j)}+|\beta
_{j}|^{2}\epsilon _{\downarrow \downarrow }^{(j)}+\alpha _{j}\beta
_{j}^{*}\epsilon _{\uparrow \downarrow }^{(j)}e^{ig_{j}t}+\alpha
_{j}^{*}\beta _{j}\epsilon _{\downarrow \uparrow }^{(j)}e^{-ig_{j}t})
\label{ast}
\end{equation}

{\bf Step 3}: Let us now compute the time evolution in the two cases.

(a) Since $\max_{t}(|\alpha _{i}|^{4}+|\beta _{i}|^{4}+2|\alpha
_{i}|^{2}|\beta _{i}|^{2}\cos 2g_{i}t)=1$ and $\min_{t}(|\alpha
_{i}|^{4}+|\beta _{i}|^{4}+2|\alpha _{i}|^{2}|\beta _{i}|^{2}\cos
2g_{i}t)=(2|\alpha _{i}|^{2}-1)^{2}$, the aleatory numbers ($|\alpha
_{i}|^{4}+|\beta _{i}|^{4}+2|\alpha _{i}|^{2}|\beta _{i}|^{2}\cos 2g_{i}t)$
fluctuate between $1$ and $(2|\alpha _{i}|^{2}-1)^{2}$. Then, from eq.(\ref
{r}) we obtain 
\[
\lim_{N\rightarrow \infty }r(t)=0 
\]
and from eq.(\ref{16}) we obtain the weak limit 
\[
\lim_{t\rightarrow \infty }\langle O_{S_{0}}\rangle _{\psi (t)}=|\alpha
_{i}|^{2}s_{00}+|\beta _{i}|^{2}s_{11}=\langle O_{S_{0}}\rangle _{\rho
_{S*}} 
\]
where 
\[
\rho _{S*}=\left( 
\begin{array}{ll}
|\alpha _{i}|^{2} & 0 \\ 
0 & |\beta _{i}|^{2}
\end{array}
\right) 
\]
This means that the proper system $S_{0}$ decoheres, according to EID, in a
finite decoherence time $t_{DS_{0}}$ (in fact, see $r(t)$ for $N=20$ and $%
N=100$ in fig.1 of \cite{Max}).

(b) But let us consider now the evolution of the spin system $S_{j}$ given
by eq.(\ref{ast}): $\langle O_{S_{j}}\rangle _{\psi (t)}$ just oscillates
and, as a consequence, it has no limit. Therefore, the generic spin system $%
S_{j}$ certainly does not decohere. This is not surprising if we recall
that, in the Hamiltonian (\ref{7}), the spin systems of the environment $%
E=\bigcup_{i}S_{i}$ are uncoupled: each spin system freely evolves and, for
this reason, the environment is unable to reach a final stable state.%
\footnote{%
They are only indirectly coupled trough $S_{0}$, but this interaction
becomes negligible when $N\rightarrow \infty $.} We could also have obtained
this result by means of SID since, as explained in Section III, if $V_{(2)}=0
$, then $t_{DU}\rightarrow \infty $.

This model shows that, by using the observables of points (a) and (b), we
gain a complete knowledge of the behavior of the system. The Hamiltonian (%
\ref{7}) is not symmetric with respect to $S_{0}$ and $S_{j}$, since $S_{0}$
is coupled to all the $S_{j}$, but the $S_{j}$ are not coupled to each other
but only coupled to $S_{0}.$ In other words, since $S_{0}$ is not trivially
coupled, it decoheres in a finite time $t_{DS_{0}}$ (see fig.1 of \cite{Max}%
), whereas the $S_{j}$ are uncoupled or ''trivially coupled'' and,
therefore, they do not decohere.

But the point to emphasize here is that we could have deduced this result
directly from Section III: the environment $E=\bigcup_{i}S_{i}$ does not
decohere (namely, it has an infinite decoherence time). As a consequence,
the whole system $U=S_{0}\cup E$ does not decohere and it would be a miracle
that it did (see figs. 2, 3 and 4 of \cite{Max}). Our general framework
allows us to understand the well known model of \cite{Max} and \cite{3A}
from a broader perspective: it can be proved that $t_{DS_{0}}\ll $ $%
t_{DU}=\infty $,\footnote{%
>From this perspective, all the criticisms to SID of paper \cite{Max}
disappear, since they were formulated from a wrong viewpoint. The model does
not show that the destructive interference of the off-diagonal terms of $%
\rho (t)$ is not efficient in all cases, as it is claimed; it merely proves
that the whole closed system does not decohere when the environment
Hamiltonian is trivial, which is a universally accepted fact. Nevertheless,
the criticisms of \cite{Max} are far from being trivial: we needed several
months to understand the puzzle.} and the results of EID and SID can be
shown to agree.

\section{Conclusion.}

The formalism sketched in this paper encompasses both EID and SID (and
probably other decoherence approaches). It is supported by the experimental
basis obtained both in EID and SID. EID has a large amount of experimental
confirmations (see \cite{Ex}). The best result of SID is the complete
description of the classical limit of quantum systems (see \cite
{CAST-GADELLA}), and the fact that the classical mechanics so obtained is
experimentally proved beyond all doubts. Moreover, according to \cite{DT},
decoherence times computed by EID and by SID are of similar order.

The new formalism supplies a solution to the main problems of the EID
approach:

i.- Closed systems do decohere and their decoherence times can be computed.

ii.- The ''looming big'' problem of EID, i.e. the fact that it does not
provide a criterion to decide where to place the cut between ''the'' proper
system $S$ and ''the'' environment $E$ in the closed system $U,$ is now
dissolved. In fact, the complete study of the system $U$ requires several
partitions $U=S_{i}\cup E_{i}.$ The choice of these partitions is arbitrary,
but the nature of each partition is given by the Hamiltonian. In fact, the
example of section IV, with Hamiltonian (\ref{7}), shows that there are only
two kinds of spin sytems to be studied: $S_{0}$ and a generic $S_{j}$. Only
once the two subsystems are studied, but not before, the whole closed system 
$U$ is completely understood.\footnote{%
This strategy was not followed in paper \cite{Max}, where the system was not
completely studied, and thus a wrong conclusion was obtained.}

iii.- The problem of rigorously defining a dynamical pointer basis (valid
for all times) via, e.g., the predictability sieve criterion, remains
unsolved. But the final pointer basis, where decoherence appears in the weak
limits, is completely well defined.

Finally, it is worth stressing that the study of the role of complexity in
decoherence presented in paper \cite{CL}, where quantum non-integrable
systems are described, may solve problems like those of paper \cite{Casati},
where quantum chaos has obviously to be invoked.

\section{Acknowledgments}

We are very grateful to Roland Omn\'{e}s and Maximilian Schlosshauer for
many comments and criticisms. This research was partially supported by
grants of the University of Buenos Aires, CONICET, and FONCYT of Argentina.


\begin{references}
\bibitem{vK}  N. G. van Kampen, {\it Physica}, {\bf 20,} 603, 1954. L. van
Hove, {\it Physica}, {\bf 23}, 441, 1957; {\it Physica}, {\bf 25}, 268,
1959. A. Daneri, A. Loinger and G. Prosperi,{\it \ Nucl. Phys}., {\bf 33, }%
297, 1962.

\bibitem{Omnes}  R. Omn\'{e}s, {\it Braz. Jour. Phys., }{\bf 35, }207, 2005.

\bibitem{Zeh}  H. D. Zeh, {\it Found. Phys.}, {\bf 1}, 69, 1970; {\it Found.
Phys.}, {\bf 3}, 109, 1973. W. H. Zurek, {\it Phys. Rev. D}, {\bf 26},
1862,1982; {\it Progr. Theor. Phys}., {\bf 89}, 281, 1993; {\it Rev. Mod.
Phys}., {\bf 75}, 715, 2003.

\bibitem{Penrose}  L. Diosi, {\it Phys. Lett. A}, {\bf 120, }377, 1987; {\it %
Phys. Rev. A,} {\bf 40, }1165, 1989. G. J. Milburn, {\it Phys. Rev. A,} {\bf %
44, }5401, 1991. R. Penrose, {\it Shadows of the Mind, }Oxford Univ. Press,
Oxford, 1995. G. Casati and B. Chirikov, {\it Phys. Rev. Lett.}, {\bf 75,}
349, 1995; {\it Physica D}, {\bf 86, }220, 1995. S. Adler, {\it Quantum
Theory as an Emergent Phenomenon, }Cambridge Univ. Press, Cambridge, 2004.

\bibitem{O'Connel}  G. W. Ford and R. F. O'Connell, ''Decoherence without
dissipation'', quant-ph/0301054, 2003.

\bibitem{PRA-2000}  M. Castagnino and R. Laura, {\it Phys. Rev. A}, {\bf 62}%
, 022107, 2000; {\it Int. Jour. Theor. Phys.,} {\bf 39}, 1767, 2000. M.
Castagnino{\it ,} {\it Physica A}, {\bf 335}, 511, 2004. M. Castagnino and
O. Lombardi{\it , } {\it Int. Jour. Theor. Phys.}, {\bf 42}, 1281, 2003; 
{\it Stud. Hist. Phil. Mod. Phys.}, {\bf 35}, 73, 2004. M. Castagnino and A.
Ordo\~{n}ez, {\it Int. Jour. Theor. Phys.}, {\bf 43}, 695, 2004.

\bibitem{Max}  M. Schlosshauer, {\it Phys. Rev. A,} {\bf 72}, 012109, 2005.

\bibitem{Omnes1}  R. Omn\'{e}s, ''Decoherence, irreversibility, and the
selection by decoherence of quantum states with definite probabilities''
arXiv-quant-ph/03041, 2003.

\bibitem{3A}  J. P. Paz and W. Zurek, ''Environment induced decoherence and
the transition from quantum to classical'', arXiv: quant-ph/0010011, 2000.

\bibitem{Ex}  E. Joos, H. D. Zeh, C. Kiefer, D. Giulini, J. Kupsch and I. O.
Stamatescu , {\it Decoherence and the Appearance of a Classical World in
Quantum Theory, }Springer Verlag, Berlin, 2003.

\bibitem{DT}  M. Castagnino, O. Lombardi, {\it Phys. Rev. A}, {\bf 72},
012102, 2005.

\bibitem{CAST-GADELLA}  M. Castagnino and M. Gadella, ''The problem of the
classical limit in quantum mechanics and the role of self-induced
decoherence'', {\it Found. Phys.}, forthcoming.

\bibitem{CL}  M. Castagnino and O. Lombardi, {\it Chaos, Solitons and
Fractals}, {\bf 28}, 879, 2006.

\bibitem{Casati}  G. Casati and T. Prosen, {\it Los} {\it Alamos Science}, 
{\bf 27}, 2, 2002.
\end{references}
\end{document}